\documentclass[aps,twocolumn,floatfix,superscriptaddress,preprintnumbers,showpacs,showkeys]{revtex4}
\usepackage{amssymb}
\usepackage{amsmath}
\usepackage{amsfonts}
\usepackage{epsfig}
\usepackage{graphicx}
\usepackage{tabularx}
\usepackage{verbatim}
\newcommand{\eq}{\begin{eqnarray}}
\newcommand{\en}{\end{eqnarray}}

\begin{document}

\title{Role of multichannel $\pi\pi$ scattering in decays of bottomia}

\author{Yurii S. Surovtsev}
\affiliation{Bogoliubov Laboratory of Theoretical Physics,
Joint Institute for Nuclear Research, 141980 Dubna, Russia}
\author{P. Byd\v{z}ovsk\'y}
\affiliation{Nuclear Physics Institute of the AS CR, 25068 \v{R}e\v{z},
Czech Republic}
\author{Thomas Gutsche}
\affiliation{Institut f\"ur Theoretische Physik,
Universit\"at T\"ubingen,
Kepler Center for Astro and Particle Physics,
Auf der Morgenstelle 14, D-72076 T\"ubingen, Germany}
\author{Robert~Kami\'nski}
\affiliation{Institute of Nuclear Physics of the PAN, Cracow 31342, Poland}
\author{Valery E. Lyubovitskij}
\affiliation{Institut f\"ur Theoretische Physik,
Universit\"at T\"ubingen,
Kepler Center for Astro and Particle Physics,
Auf der Morgenstelle 14, D-72076 T\"ubingen, Germany}
\affiliation{Department of Physics, Tomsk State University,
634050 Tomsk, Russia}
\affiliation{Mathematical Physics Department,
Tomsk Polytechnic University,
Lenin Avenue 30, 634050 Tomsk, Russia}
\author{Miroslav Nagy}
\affiliation{Institute of Physics, SAS, Bratislava 84511, Slovak Republic}

\date{\today}

\begin{abstract}
The effect of isoscalar S-wave multichannel $\pi\pi\to\pi\pi,K\overline{K},\eta\eta$ scattering is 
considered in the analysis of decay data of the $\Upsilon$-mesons. We show that when allowing for 
the final state interaction contribution to the decays $\Upsilon(mS)\to\Upsilon(nS)\pi\pi$ 
($m>n, m=2,3, n=1,2$) in our model-independent approach, we can explain the two-pion
energetic spectra of these $\Upsilon$ transitions including the two-humped shape of the di-pion mass 
distribution in  $\Upsilon(3S)\to\Upsilon(1S)\pi\pi$ as the coupled-channel effect. 
It is shown also that the considered bottomia decay data do not offer new insights into the nature 
of the $f_0$ mesons, which were not already deduced in our previous analyses of pseudoscalar 
meson scattering data.
\end{abstract}



\maketitle

\section{Introduction} 

A comprehensive study of the properties of scalar mesons is important for the most profound 
topics concerning the QCD vacuum, because both sectors affect each other due to possible 
``direct" transitions between them. The problem of a unique structure interpretation of the 
scalar mesons is far away from being solved completely~\cite{PDG-14}. 
For example, applying our model-independent method in the three-channel analyses of processes 
$\pi\pi \to \pi\pi, K\overline{K}, \eta\eta, \eta\eta^\prime$~\cite{SBL-prd12} 
we have obtained parameters for the scalar mesons $f_0(500)$ and $f_0(1500)$, 
which considerably differ from results of analyses which utilize other methods --- 
mainly those based on dispersion relations and Breit--Wigner approaches 
(see detailed discussion in Ref.~\cite{SBL-prd12}).

The decays of heavy quarkonia into a pair of pseudoscalar mesons and 
a spectator are a good laboratory for studying the $f_0$ mesons 
if the pseudoscalar meson pair is produced in an $S$-wave. 
This occurs, e.g., in the specific decays of charmonia and bottomia:
$J/\psi\to\phi(\pi\pi, K\overline{K})$, $\psi(2S)\to J/\psi\,\pi\pi$
and $\Upsilon(2S)\to\Upsilon(1S)\pi\pi$,
$\Upsilon(3S)\to\Upsilon(1S)\pi\pi$,
$\Upsilon(3S)\to\Upsilon(2S)\pi\pi$.

In recent papers~\cite{SBGLKN-npbps13,SBLKN-prd14} we have already
presented a combined analysis of the isoscalar $S$-wave processes
$\pi\pi\to\pi\pi,K\overline{K},\eta\eta$. The analysis was performed
in our model-independent approach based on analyticity, unitarity, 
on the use of the uniformization procedure, and on the inclusion of
charmonium decay processes
$J/\psi\to\phi(\pi\pi, K\overline{K})$, $\psi(2S)\to J/\psi\,\pi\pi$.
Taking into account the data on charmonium decays helped to narrow down 
the $f_0(500)$ solution to the one with the larger width. Other resonance parameters 
were practically not changed after using the charmonium data. 
At this stage it is worth performing a combined analysis including data on decays 
of the $\Upsilon$-meson family: $\Upsilon(2S)\to\Upsilon(1S)\pi\pi$, 
$\Upsilon(3S)\to\Upsilon(1S)\pi\pi$ and $\Upsilon(3S)\to\Upsilon(2S)\pi\pi$. 
These decays have been studied intensively using various approaches 
(see, e.g., Ref.~\cite{SimVes} and the references therein).

Note that here, except for a possible confirmation and specification of the scalar 
resonance parameters, there is the problem of explaining the two-humped shape of 
the di-pion mass distribution in the decay $\Upsilon(3S)\to\Upsilon(1S)\pi\pi$. 
This distribution might be the result of the destructive interference of the relevant 
contributions to the decay $\Upsilon(3S)\to\Upsilon(1S)\pi\pi$. 
However, in this scenario the phase space cuts off possible contributions, 
which might interfere destructively with the $\pi\pi$ scattering contribution 
giving the specific shape of the di-pion spectrum.

After the experimental evidence for the two-humped shape of the di-pion spectrum 
Lipkin and Tuan~\cite{Lipkin-Tuan} suggested that the decay $\Upsilon(3S)\to\Upsilon(1S)\pi\pi$ 
proceeds as follows: $~\Upsilon(3S)\to B\overline{B}\to B^\ast\overline{B}\pi\to
B\overline{B}\pi\pi\to\Upsilon(1S)\pi\pi$.
In the heavy-quarkonium limit, when neglecting the recoil of the
final quarkonium state, they obtained a transition amplitude containing a term proportional to 
${{\bf p}_1\!\cdot{\bf p}_2}\propto\cos\theta_{12}$, where $\theta_{12}$ is the angle between 
the pion three-momenta ${\bf p}_1$ and ${\bf p}_2$, multiplied by some function of 
the kinematical invariants. If the latter was a constant, then the angular distribution
$d\Gamma/d\cos\theta_{12}\propto\cos\theta_{12}^2$
(and $d\Gamma/d M_{\pi\pi}$) would have the two-humped shape.
However, this scenario was not tested numerically by fitting to data. 
It is possible that this effect is negligible due to the small coupling of $\Upsilon$ 
to the b-flavored sector.

In Ref.~\cite{Moxhay} Moxhay suggested that the two-humped shape is
a result of the interference between two parts of the decay amplitude. 
The first part, in which the $\pi\pi$ final state interaction is allowed for, 
is related to a mechanism which acts as well in the decays of excited quarkonia states 
$\psi(2S)\to J/\psi\,\pi\pi$ and $\Upsilon(2S)\to\Upsilon(1S)\pi\pi$ and which, 
obviously, should also occur in the process $\Upsilon(3S)\to\Upsilon(1S)\pi\pi$. 
The second  part is responsible for the Lipkin--Tuan mechanism. 
However, nothing remains from the latter term because the author says that this part does 
not dominate the amplitude and ``the other tensor structures conspire to give a distribution 
in $M_{\pi\pi}$ that is more or less flat'' -- it is constant. 
It seems that the approach of Ref.~\cite{Komada-Ishida-Ishida} resembles the above one. 
The authors simply supposed that a pion pair is formed in the $\Upsilon(3S)$ decay both 
as a result of rescattering and direct production.
One can, however, believe that the latter is not reasonable because the pions interact strongly.
In the present paper we show that the indicated effect of destructive
interference can be achieved by taking into account our previous
conclusions on the wide resonances~\cite{SBLKN-prd14,SBLKN-jpgnpp14}, 
without any further assumptions.

\section{Multichannel $\pi\pi$ scattering in the decays of bottomia}

When carrying out our combined analysis, data for the processes
$\pi\pi\to\pi\pi,K\overline{K},\eta\eta$ were taken from many sources 
(see the corresponding references in~\cite{SBLKN-prd14}).
For the $J/\psi\to\phi\pi\pi,\phi K\overline{K}$ decays data were taken 
from the Mark III, DM2 and BES II collaborations;
for $\psi(2S)\to J/\psi(\pi^+\pi^-)$ --- from Mark~II;
for $\psi(2S)\to J/\psi(\pi^0\pi^0)$ --- from Crystal Ball(80)
(see corresponding references also in~\cite{SBLKN-prd14}).
For $\Upsilon(2S)\to\Upsilon(1S)(\pi^+\pi^-,\pi^0\pi^0)$ data were taken
from ARGUS~\cite{Argus}, CLEO~\cite{CLEO,CLEO07}, CUSB~\cite{CUSB},
and the Crystal Ball~\cite{Crystal_Ball(85)} collaborations.
Finally, for $\Upsilon(3S)\to\Upsilon(1S)(\pi^+\pi^-,\pi^0\pi^0)$ and
$\Upsilon(3S)\to\Upsilon(2S)(\pi^+\pi^-,\pi^0\pi^0)$ 
measurements are available from the CLEO collaboration~\cite{CLEO(94),CLEO07}.

The formalism for calculating dimeson mass distributions in the decays 
$J/\psi\to\phi(\pi\pi, K\overline{K})$ and
$V^{\prime}\to V\pi\pi$ ($V=\psi,\Upsilon$) can be found in
Ref.~\cite{MP-prd93}. It was assumed that the pairs of pseudoscalar mesons 
in the final state have zero isospin and spin. Only these pairs of pseudoscalar 
mesons undergo final state interactions, whereas the final vector meson ($\phi$, $V$) 
acts as a spectator. The amplitudes for the decays include the scattering amplitudes 
$T_{ij}$ $(i,j=1-\pi\pi,2-K\overline{K})$ as follows: 
\begin{equation}
F_n(s) = (\rho_{n0}+\rho_{n1}\,s)\,T_{11} +(\omega_{n0}+\omega_{n1}\,s)\,T_{21},
\end{equation}
where $n=$ 1, 2, and 3 denotes the considered decays
$\Upsilon(2S)\to \Upsilon(1S)\pi\pi$, $\Upsilon(3S)\to \Upsilon(1S)\pi\pi$,
and $\Upsilon(3S)\to \Upsilon(2S)\pi\pi$, respectively.
The free parameters $\rho_{n0}$, $\rho_{n1}$, $\omega_{n0}$, and $\omega_{n1}$
depend on couplings of $\Upsilon(2S)$ and $\Upsilon(3S)$
to the channels $\pi\pi$ and $K\overline{K}$.
The amplitudes $T_{ij}$ are expressed through the $S$-matrix elements
\eq
S_{ij}=\delta_{ij}+2i\sqrt{\rho_1\rho_2}T_{ij}
\en
where $\rho_i=\sqrt{1-s_i/s}$ and $s_i$ is the reaction threshold. 
The $S$-matrix elements are parametrized on the uniformization plane of the 
$\pi\pi$ scattering amplitude by poles and zeros which represent resonances. 
The uniformization plane is obtained by a conformal map of the 8-sheeted Riemann surface, 
on which the three-channel $S$ matrix is determined, onto the plane. In the uniformizing 
variable used we have neglected the $\pi\pi$-threshold branch point and allowed for the 
$K\overline{K}$- and $\eta\eta$-threshold branch points and left-hand branch point at $s=0$ 
related to the crossed channels. The background is introduced to the amplitudes in a natural way: 
on the threshold of each important channel there appears generally speaking a complex phase shift. 
It is important that we have obtained practically zero background of the $\pi\pi$ scattering 
in the scalar-isoscalar channel. It confirms well our representation of resonances.

The expressions for the decay $\Upsilon(2S)\to\Upsilon(1S)\pi\pi$
\eq
N|F|^{2}\sqrt{(s-s_1) \lambda(m_{\Upsilon(2S)}^2,s,m_{\Upsilon(1S)}^2)}\,,
\en
where $\lambda(x,y,z)=x^2+y^2+z^2-2xy-2yz-2xz$ is the K\"all\'en function, 
and the analogue relations for $\Upsilon(3S)\to \Upsilon(1S,2S)\pi\pi$ give the dimeson 
mass distributions. $N$ (normalization to experiment) is as follows: 
for $\Upsilon(2S)\to \Upsilon(1S)\pi^+\pi^-$, 4.3439 for ARGUS, 2.1776 for CLEO(94), 
1.2011 for CUSB; for $\Upsilon(2S)\to\Upsilon(1S)\pi^0\pi^0$, 0.0788 for Crystal Ball(85); 
for $\Upsilon(3S)\to\Upsilon(1S)(\pi^+\pi^-~{\rm and}~\pi^0\pi^0)$,  
0.5096 and 0.2235 for CLEO(07), 
and for $\Upsilon(3S)\to\Upsilon(2S)(\pi^+\pi^-$ ${\rm and}~\pi^0\pi^0)$,
7.7397 and 3.8587 for CLEO(94), respectively. 
The parameters of the coupling functions of the decay particles 
[$\Upsilon(2S)$ and $\Upsilon(3S)$] to channel~$i$, obtained in the analysis, are 
$(\rho_{10},\rho_{11},\omega_{10},\omega_{11})=$
(0.4050, 47.0963, 1.3352,$-21.4343)$,
$(\rho_{20},\rho_{21},\omega_{20},\omega_{21})
=$($1.0827,-2.7546$,$0.8615$,~0.6600),
$(\rho_{30},\rho_{31},\omega_{30},\omega_{31})=(7.3875,-2.5598,0.0,0.0)$.

A satisfactory combined description of all considered processes is obtained with 
a total $\chi^2/\mbox{ndf}=640.302/(564-70)\approx1.30$; for the $\pi\pi$ scattering, 
$\chi^2/\mbox{ndf}\approx1.15$; for
$\pi\pi\to K\overline{K}$, $\chi^2/\mbox{ndf}\approx1.65$;
for $\pi\pi\to\eta\eta$, $\chi^2/\mbox{ndp}\approx0.87$;
for decays $J/\psi\to\phi(\pi^+\pi^-, K^+K^-)$,
$\chi^2/\mbox{ndp}\approx1.21$;
for $\psi(2S)\to J/\psi(\pi^+\pi^-,\pi^0\pi^0)$,
$\chi^2/\mbox{ndp}\approx2.43$;
for $\Upsilon(2S)\to\Upsilon(1S)(\pi^+\pi^-,\pi^0\pi^0)$,
$\chi^2/\mbox{ndp}\approx1.01$;
for $\Upsilon(3S)\to\Upsilon(1S)(\pi^+\pi^-,\pi^0\pi^0)$,
$\chi^2/\mbox{ndp}\approx0.97$; 
for $\Upsilon(3S)\to\Upsilon(2S)(\pi^+\pi^-,\pi^0\pi^0)$,
$\chi^2/\mbox{ndp}\approx0.54$.

In Figs.~1 and 2 we show the fits to the experimental data on above
indicated bottomia decays in the combined analysis with the processes 
$\pi\pi\to\pi\pi,K\overline{K},\eta\eta$ and the decays
$J/\psi\to\phi\pi\pi,\phi K\overline{K}$. The dips in the energy
dependence of di-pion spectra (Fig.~2, upper panel) are the result of 
a destructive interference between the $\pi\pi$ scattering and
$K\overline{K}\to\pi\pi$ contributions to the final states of 
the decays $\Upsilon(3S)\to\Upsilon(1S)(\pi^+\pi^-,\pi^0\pi^0)$.

The description of the processes $\pi\pi\to\pi\pi,K\overline{K},\eta\eta$ and charmonia decays 
and the resulting resonance parameters practically did not change
when compared to the case without bottomia decays. The description of the respective data 
and the resonance parameters can be found in Refs.~\cite{SBGLKN-npbps13,SBLKN-prd14}.

\section{Summary}

The combined analysis was performed for data on isoscalar S-wave processes 
$\pi\pi\to\pi\pi,K\overline{K},\eta\eta$ and on the decays of heavy quarkonia 
$J/\psi\to\phi(\pi\pi, K\overline{K})$, $\psi(2S)\to J/\psi\,\pi\pi$, 
$\Upsilon(2S)\to\Upsilon(1S)\pi\pi$, $\Upsilon(3S)\to\Upsilon(1S)\pi\pi$ and 
$\Upsilon(3S)\to\Upsilon(2S)\pi\pi$ from the ARGUS, Crystal Ball,
CLEO, CUSB, DM2, Mark~II, Mark~III, and BES~II collaborations.
It was shown that in the final states of the bottomia decays 
the contribution of the coupled processes, e.g., $K\overline{K}\to\pi\pi$, 
is important even if these processes are energetically forbidden. 
This is in accordance with our previous conclusions on wide 
resonances~\cite{SBLKN-prd14,SBLKN-jpgnpp14,SBKLN-PRD12}: 
when a wide resonance cannot decay into a channel, which opens above 
its pole mass and which is strongly coupled [e.g. the $f_0(500)$ and the
$K\overline{K}$ channel], one should consider this resonance as
a multichannel state. E.g., on the basis of this consideration the new and 
natural mechanism of destructive interference in the decay
$\Upsilon(3S)\to\Upsilon(1S)\pi\pi$ is indicated, which provides the
two-humped shape of the di-pion mass distribution (Fig.~2).

The results of the analysis confirm all of our earlier conclusions
on the scalar mesons~\cite{SBLKN-prd14}. Hence the considered bottomia decay data do not offer new insights into the nature of the scalar mesons, which were not already deduced in previous analyses of pseudoscalar meson scattering data.

\begin{acknowledgments} 

This work was supported in part by the Heisenberg-Landau Program,
the Votruba-Blokhintsev Program for Cooperation of Czech Republic
with JINR, the Grant Agency of the Czech Republic (Grant No. P203/12/2126), 
the Bogoliubov-Infeld Program for Cooperation of Poland with JINR, 
by the Grant Program of Plenipotentiary of Slovak Republic at JINR, 
the DFG under Contract No. LY 114/2-1, 
the Tomsk State University Competitiveness Improvement Program, and 
by the Polish National Science Center (NCN) Grant No. DEC-2013/09/B/ST2/04382.

\end{acknowledgments}

\clearpage

\begin{widetext}
\begin{figure}[htb]
\begin{center}
\epsfig{figure=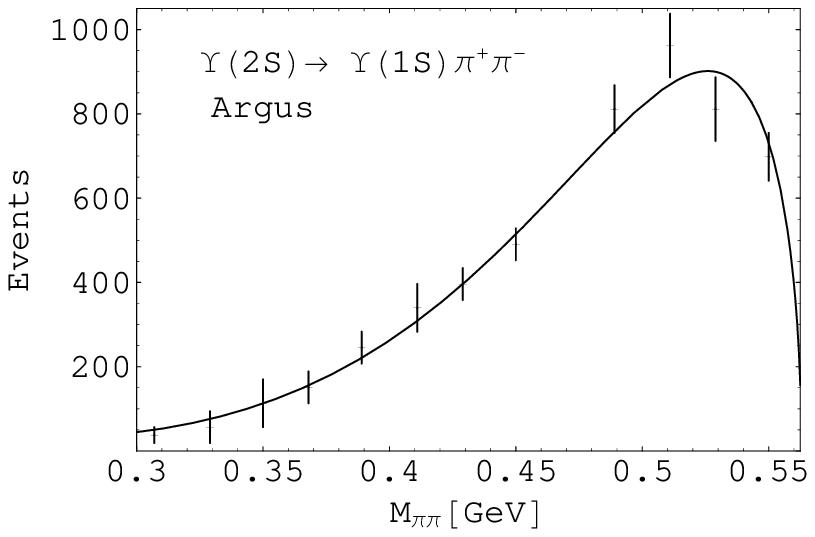,scale=.9}\hspace*{.25cm}
\epsfig{figure=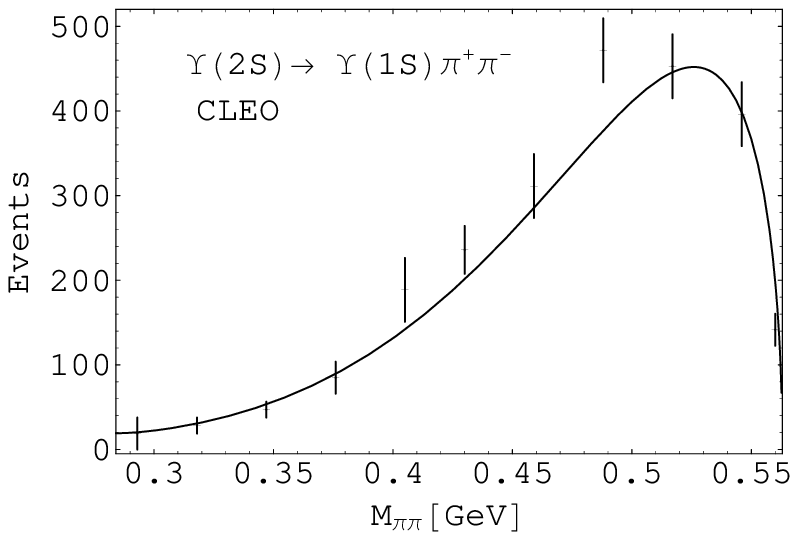,scale=.9}\\
\epsfig{figure=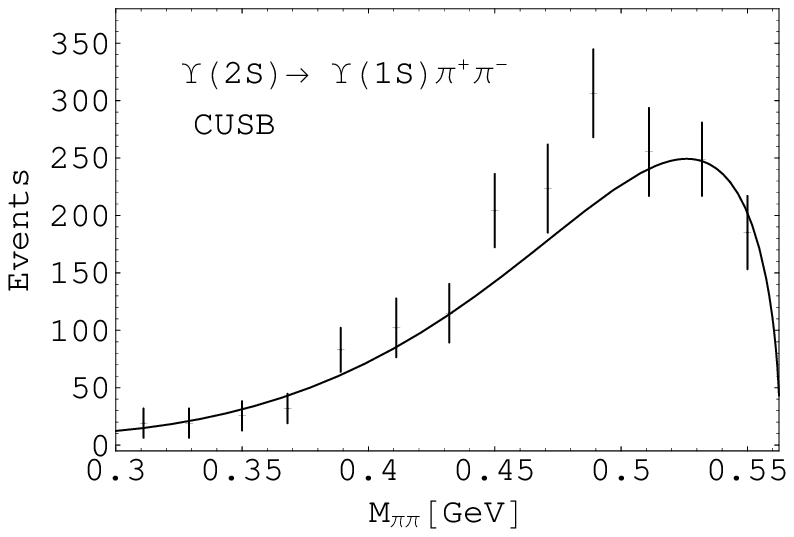,scale=.9}\hspace*{.25cm}
\epsfig{figure=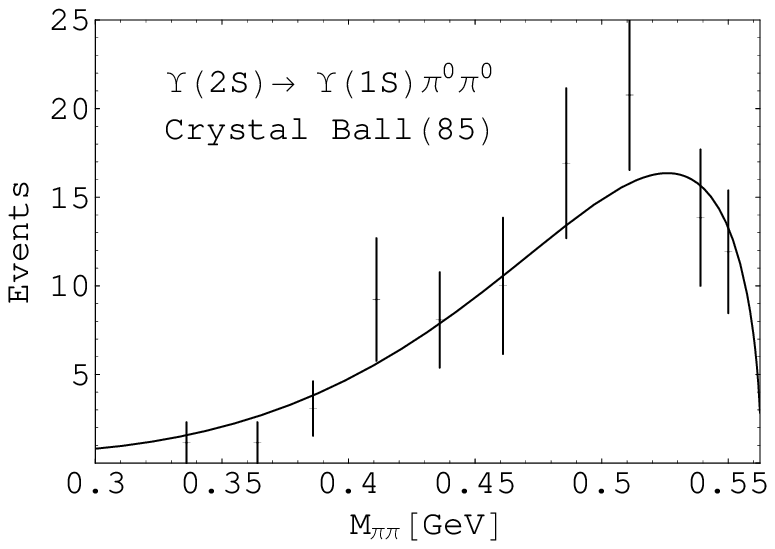,scale=.9}
\caption{Decay $\Upsilon(2S)\to\Upsilon(1S)\pi\pi$.
\label{fig:Ups21}}

\vspace*{.5cm }
\epsfig{figure=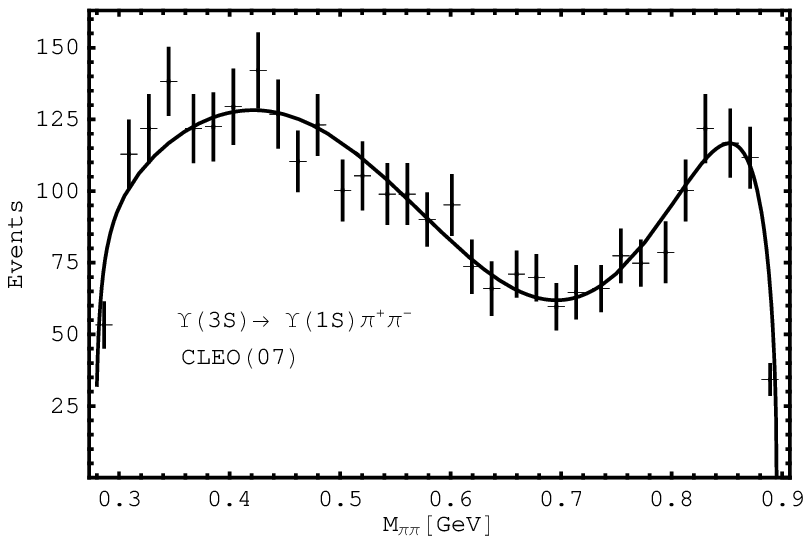,scale=.87}\hspace*{.25cm}
\epsfig{figure=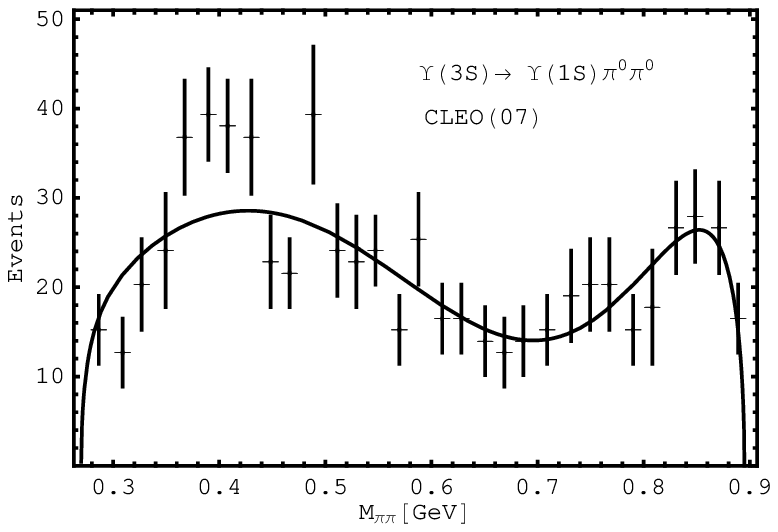,scale=.87}\\
\epsfig{figure=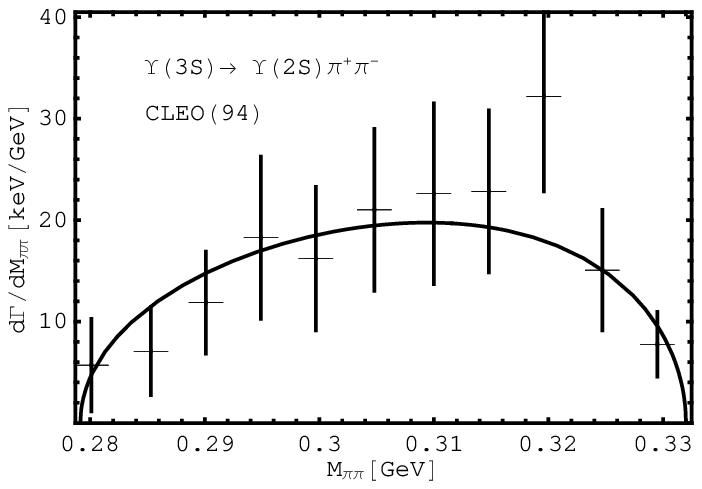,scale=.9}\hspace*{.25cm}
\epsfig{figure=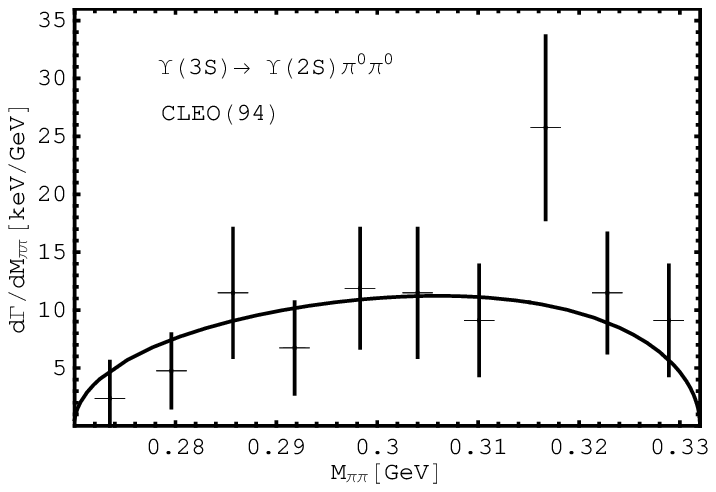,scale=.9}
\vspace*{-1cm }
\caption{Decays $\Upsilon(3S)\to\Upsilon(nS)\pi\pi$\,, \ $n=1,2$.
\label{fig:Ups31_32}}
\end{center}
\end{figure}

\end{widetext}

\end{document}